# 320-Channel Dual Phase Lock-in Optical Spectrometer

P. S. Fodor, S. Rothenberger, and J. Jevy

Department of Physics and Astronomy, University of Pittsburgh, Pittsburgh, PA 15260

**Abstract:**

The development of a multiple-channel lock-in optical spectrometer (LIOS) is presented, which enables parallel phase-sensitive detection at the output of an optical spectrometer. The light intensity from a spectrally broad source is modulated at the reference frequency, and focused into a high-resolution imaging spectrometer. The height at which the light enters the spectrometer is controlled by an acousto-optic deflector, and the height information is preserved at the output focal plane. A two-dimensional InGaAs focal plane array collects light that has been dispersed in wavelength along the horizontal direction, and in time along the vertical direction. The data is demodulated using a high performance computer-based digital signal processor. This parallel approach greatly enhances (by more than 100x) the speed at which spectrally resolved lock-in data can be acquired. The noise performance of a working system optimized for the 1.3 $\mu$m wavelength range is analyzed using a laser diode light source. Time-resolved absorption traces are obtained for InAs quantum dots embedded in a GaAs matrix, and for dispersed films of PbSe nanocrystals.

# I. INTRODUCTION

Optical spectroscopy is a pervasive tool across physical, chemical and biological sciences. With a typical spectrometer, light focused on the entrance slit is dispersed by a diffraction grating which spatially separates the light according to wavelength. Spectroscopic information is obtained by scanning the angle of the diffraction grating, allowing different wavelengths to pass through an exit slit. The availability of charge-coupled device (CCD) and focal plane array (FPA) detectors has greatly reduced the amount of time required for the acquisition of optical spectra. Further parallelization has been achieved by using so-called "imaging" spectrometers that are optimized to preserve the vertical point-spread function of the spectrometer, allowing multiple input fibers to be imaged onto different horizontal portions of the detector array without interference[1]. The increased throughput of such parallel techniques has become critical for applications such as Raman spectroscopy[2] and near-field scanning optical microscopy[3], where the signals of interest are small and background-free.

For many applications, the signal of interest is superimposed upon a large noisy background. Applications that fit this description include spectrophotometry, and a wide variety of linear and nonlinear spectroscopies in which information about the sample is encoded in the intensity or polarization of light.[4,5] A simple background subtraction can be achieved by synchronizing the frame acquisition to an optical chopper or similar modulator. However, the spectral response of such a technique is functionally equivalent to a comb filter and thus is less effective than bandwidth narrowing methods that use true lock-in amplification[6,7]. Furthermore, the range of modulation frequencies is limited by the camera frame rate, which is usually quite low. Laser noise usually falls off significantly above a few kHz, higher than the frame rates of most high-sensitivity photometric imaging devices like CCDs and FPAs. As a consequence, true phase-sensitive detection or lock-in amplification[8-10] is the method of choice in the above mentioned situations, when modulation at frequencies higher than a few kHz is desired. However, due to the high photon fluxes and large backgrounds, it is not straightforward to parallelize lock-in techniques.

Here we present an implementation of a multiple-channel lock-in optical spectrometer (LIOS) which enables phase sensitive detection methods to be applied in parallel, greatly enhancing the speed (by more than 100x) with which optical spectra can be acquired. The operating principle of LIOS is described, and operation is demonstrated using a commercial 1310 nm laser diode. Also, time-resolved absorption measurements are presented for InAs quantum dots embedded in a GaAs matrix, as well as for dispersed films containing PbSe nanocrystals.

## II. PRINCIPLE OF OPERATION

### A. Overview

A straightforward implementation of a LIOS requires a one-dimensional detector array and a minimum of one lock-in channel per detector. Lock-in amplifiers that employ digital-signal-processing (DSP)[11] methods have rapidly replaced the original analog lock-in amplifier design invented by Dicke[7]. Digital lock-ins have superior phase stability and dynamic range, making them favorable for most demanding applications except perhaps milikelvin transport experiments where digital circuitry is usually avoided. Banks of up to 32 digital lock-in channels are commercially available[12], but for a reasonable number of channels (*e.g.,* 256 or 512), this approach quickly becomes financially burdensome and unwieldy.

Here an alternative approach is described that combines the parallelism of an imaging device with the DSP capabilities of modern lock-in amplifiers. The optical diagram of LIOS is depicted in Figure 1. The key elements are a high-efficiency acousto-optic deflector (AOD) (*A-A Opto-Electronic*), a high-resolution imaging spectrometer (*McPherson Inc.*), a two-dimensional InGaAs FPA (Phoenix NIR Camera, *Indigo Systems*) and a high-performance computer-based DSP. The FPA collects spectral information along the horizontal direction (dispersed by the spectrometer), and temporal information along the vertical direction (dispersed by the AOD). The time-domain information is Fourier-decomposed using a computer-based DSP to produce X and Y lock-in outputs for every wavelength channel. A more detailed description of the operation of LIOS is given below.

Light from the experiment is coupled into a polarization-preserving single-mode optical fiber, and expanded and collimated into a 6 mm diameter beam. The AOD chosen for the LIOS has an efficiency of 85% at 1.3um, a deflection range of 40 mrad and a bandwidth of 30 MHz at 1000 nm. Generally, better vertical imaging is achieved with longer focal length $F$ spectrometers, so a $F$=2.0 m spectrometer (f/#=14.1) is chosen. A ray-tracing analysis of the spectrometer indicates that a 10 μm diameter spot incident on the entrance slit forms an approximately 32 μm spot at the detector plane.

The (FPA) consists of a two-dimensional array (320 horizontal x256 vertical) of InGaAs square pixel detectors, 30 μm in size, with sample-and-hold circuitry behind each pixel. The maximum diffraction efficiency of the AOD covers only about half of the vertical range of the camera, so a subset of the camera, consisting of $n_v$=128 vertical pixels x $n_h$=320 horizontal pixels, is used. The FPA is sensitive in the near-infrared region (0.8 µm-2.6 µm) and has a specified noise-equivalent irradiance <1.5x$10^{-7}$ W cm$^{-2}$ for the gain setting employed. The camera is equipped with four 14-bit analog-to-digital converters operating from a 40MHz bus, allowing data transfer rates of up to 80MB/sec. Optimal throughput is achieved using "integrate-while-read" mode, in which data from the previous frame is transmitted while the current frame is being acquired.

The AOD diffracts the light, scanning it vertically along the entrance slit of the spectrometer. Thus, at the exit port the light is dispersed by the high resolution imaging spectrometer along the horizontal axis according to wavelength, and by the AOD along the vertical direction according to time (Figure 1). The AOD is controlled by a radio-frequency (RF) source that can be amplitude-modulated (AM) and frequency-modulated (FM) to control the intensity and angle of the diffracted beam, respectively. Both AM and FM control signals are produced by synchronized arbitrary waveform generators. The encoded optical information in both the spectral and time domains, is collected by the FPA. The acquisition is triggered by a TTL signal that is phase-locked to the reference frequency $f_R$, enabling phase-sensitive detection. Subsequently, the data is demodulated using a high performance computer-based digital signal processor.

The light deflection methods are not limited to acousto-optic deflection (Figure 2(a)). For broadband spectral applications, an AOD provides 50-70% diffraction efficiency. Compared to a scanned mirror, this efficiency represents a significant degradation of the signal. Scanned mirrors have their own difficulties. To achieve the large scan angles necessary for this application, either a resonant mirror scanner (Figure 2(b)) or polygon scanner (Figure 2(c)) would be required. The high quality factor and narrow frequency tunability of resonant scanners make them unattractive for phase-sensitive applications. Furthermore, given the sinusoidal nature of their deflection angle, the amount of "dead time" becomes effectively comparable to the diffraction efficiency of an AOD. Polygon scanners do not have the same problems with dead time, but are themselves difficult to phase-lock to external sources in the stable manner required by phase sensitive methods. Consequently, the flexibility and programmability of the AOD far outweigh its limitations.

Another advantage of the AOD method is the ability to use the AM capabilities for calibrating the LIOS. The calibration procedure will be described below.

### B. Digital Signal Processing

The sequence of images obtained by the FPA is transferred in real time to a PC (Pentium Xeon dual processor @ 2.2GHz, 4 GB RAM) which performs the DSP tasks. To discuss the signal processing, we restrict ourselves to the case of monochromatic light, so that only a single spectral channel is involved. The generalization to multiple channels is straightforward. We adopt the convention that functions of time (treated as a continuous index *t*) are represented with tildes (*e.g.*, $\tilde{f}(t)$), and occasionally are written without an explicit time index (*i.e.*, $\tilde{f}$). In a LIOS measurement, the signal of interest is modulated at a fixed reference frequency $f_R = 1/T_R$. The signal $\tilde{S} \equiv \{\tilde{S}(t), t_0 \leq t < t_0 + mT_R\}$, where *m* is an integer, represents the time-dependent intensity measured by a fixed detector. This signal is encoded by LIOS in the following manner. As seen in Figure 3, the frequency of the AOD is ramped linearly in time, triggered by the phase of the reference signal. The diffracted light from the AOD is focused by a lens into the spectrometer and subsequently imaged onto a one-dimensional vertical array of pixels from the FPA. The signal image is obtained as the difference of two separate frames A and B obtained after triggering on the positive and negative edge of the reference signal. This provides a straightforward reduction of the background arising from the dark count of the camera detector and from the unmodulated scattered light, allowing the extraction of small relevant signals (Figure 4).

In principle, it should be possible to calibrate the spectrometer in order to extract lock-in information from the measured array. While it is possible to model the entire system accurately, taking into account nonlinearities and non-uniformities from the AOD, and variations in pixel gain and offsets from the FPA, there is a much more straightforward and accurate method of extracting lock-in signals from the measured response.

The LIOS detector can be regarded as a linear mapping $C$ from the input signal $\tilde{S} = \tilde{S}(t)$ onto an array of pixel intensities $S = \{S_k, k = 1,...,n_V\}$: $\tilde{S} \xrightarrow{C} S$. Because there are a finite number of pixels, this mapping is not invertible; however, it preserves the spectral information of interest, namely, the Fourier components at the reference frequency as well as higher harmonics, provided the number of effective vertical pixels exceeds the Nyquist criterion: $n_V^{eff} \geq 2m$. (The number of effective vertical pixels per reference period T is given by $n_v^{eff} = n_v \cdot \frac{t_{scan} f_R}{PSF}$, where $t_{scan}$ is the time it takes to scan the FPA with the AOD, and PSF is the point-spread function FWHM. For the present system, the PSF ~ 3 with a 10 x 10 μm aperture).

For lock-in detection, it is necessary to define an appropriate scalar product or metric for applying Fourier decomposition methods to the image of the original signal. The following metric is useful for determining the Fourier coefficients of a signal $\tilde{S}(t)$ that is periodic in $T_R$: $\langle \tilde{A} | \tilde{B} \rangle \equiv \frac{2}{T_R} \int_0^{T_R} \tilde{A}(t)\tilde{B}(t) dt$. With this definition, a natural basis for the space of square-integrable functions is given by $\tilde{X}_0(t) \equiv 1/\sqrt{2}$, $\tilde{X}_k(t) \equiv \cos(2\pi k f_R t)$, $k = 1, 2, ...$ and $\tilde{Y}_l(t) \equiv \sin(2\pi l f_R t)$, $l = 1, 2, ...$. These basis functions have the property $\langle \tilde{X}_k | \tilde{Y}_l \rangle = 0$ and $\langle \tilde{X}_k | \tilde{X}_l \rangle = \langle \tilde{Y}_k | \tilde{Y}_l \rangle = \delta_{kl}$, where $\delta_{kl}$ is the Kronecker delta function. Fourier decomposition of an arbitrary function $\tilde{S}(t) = \sum_{k=0}^{\infty} a_k \tilde{X}_k(t) + \sum_{k=1}^{\infty} b_k \tilde{Y}_k(t)$ is given by $a_n = \langle \tilde{S} | \tilde{X}_n \rangle$ and $b_n = \langle \tilde{S} | \tilde{Y}_n \rangle$.

Because there is no direct access to $\tilde{S}(t)$, the task is to find the best estimate for $a_n$ and $b_n$, given the measured array of pixel values $S_k$. Formally, this can be done by defining a scalar product for the domain of **C**: $[A, B] \equiv \langle \mathbf{C}^{-1}\tilde{A} | \mathbf{C}^{-1}\tilde{B} \rangle$. The mapping **C** is singular (non-invertible), but singular value decomposition (SVD) may be used to find the best estimates for $a_n$ and $b_n$.

The first step is to calibrate the system by measuring images of the basis functions $\tilde{X}_n$ and $\tilde{Y}_n$. These functions are measured by amplitude modulating the AOD at the appropriate frequency and phase. A matrix $C$ can be constructed as follows: $C = (X_0 | X_1 | Y_1 | X_2 | Y_2 | \cdots)^T$. This matrix is a representation of the mapping **C**, and can be used construct a metric. Using SVD, we may write $C = U \cdot \Lambda \cdot V^T$. A metric $M$ can be generated from this decomposition, with the property that: $C^T \cdot M \cdot C = I$. Simple mathematics gives: $M = U \cdot \Lambda^{-2} \cdot U^T$. The determination of M for a single wavelength channel is graphically represented in Fig. 5. An identical algorithm is used to determine the corresponding metrics for all the wavelength channels employed. The calibration procedure must be repeated if the modulation frequency or the central wavelength (grating position) are changed, or if the overall light intensity (not the ac modulation) varies significantly. Once the M and C matrices are known for all the wavelength channels, the Fourier coefficients of interest $ab \equiv (a_0, a_1, b_1, a_2, b_2, ...)$ can be quickly calculated from a measurement $S$ using the metric $M$ as follows: $ab = [C, S] = C^T \cdot M \cdot S$. Typically, the components of the signal are calculated up to the second harmonic. However, if desired higher harmonics can be easily included, with an associated increase in computational resources. The range of available frequencies $f_R$ is limited only by the bandwidth of the AOD, which in our case is a few MHz. For high frequencies, a "multiple-coat" method is used in which the beam is scanned vertically an integer number of times across the FPA, before the data is transferred to the DSP.

Figure 6 shows sine and cosine reference images obtained by acquiring 1000 individual frames for a modulation frequency $f_R$ = 42 kHz. The 5º shear observed in the images comes from intrinsic dispersion in the AOD. This shear does not impact in any way the performance of the design, since each wavelength channel is demodulated using its own unique metric, independent of the other channels. Vertical cuts along the wavelength channels give periodic functions as expected. The number of periods observed is determined by $t_{scan} f_R$, where $t_{scan}$ is the vertical scanning time.

The functionality of the lock-in spectrometer has been tested using a 1310 nm diode laser (*Thorlabs* Inc) with the intensity modulated at 42 kHz. After demodulation of the acquired signal image, a high-resolution spectrum (8 nm in range) of the diode output is obtained (Figure 7). The speed at which spectrally resolved data can be acquired is enhanced by more than two orders of magnitude, compared with a scanned diffraction grating setup.

**C. Noise characterization**

The system noise is characterized using an optical parametric oscillator (*Spectra Physics Opal*), tuned at 1250 nm with a FWHM of 17 nm. In particular, the RMS noise was measured as a function of the input intensity. Typically, three major noise sources have to be considered: camera noise (itself comprised of thermal noise and electronic circuit noise), laser noise and shot noise[13]. The different noise sources can be distinguished by measuring their different dependence on the light intensity. While the camera noise is independent on the intensity, both laser and shot noise monotonically increase with the intensity following different dependencies: linear and square root, respectively. For the present setup, the noise data is accurately fitted with a linear function added to a constant background (figure 8), indicating that the performance of the system is limited only by the laser noise and the intrinsic noise of the camera, with the former becoming dominant above 1000 counts/wavelength channel.

## III. EXPERIMENTAL RESULTS

The implemented system, optimized for 1.3 μm wavelengths, was used to study carrier dynamics in two semiconductor systems: dispersed films of PbSe nanocrystals prepared through an organo-metallic route (*Evident Technologies*) and InAs quantum dot arrays grown epitaxially on GaAs substrates. Both systems are optically active in the 1.2 - 1.5 μm range in which the efficiency of the lock-in spectrometer peaks.

Time-resolved signals are obtained from the samples using a pump-probe experiment in which ultrashort pump pulses (120 fs) create well-defined carrier populations in the sample, which are subsequently probed using time-delayed probe pulses. A typical optical set-up employed in the experiment (Figure 9) allows both degenerate ($\lambda_{pump} = \lambda_{probe}$) and non-degenerate ($\lambda_{pump} < \lambda_{probe}$) experiments to be performed. A mechanical optical delay stage with a retro-reflector is used to change the delay between the pump and probe beam. For the setup presented, the time resolution is not limited by the delay stage, but by the pulse stretcher used to expand the probe pulse from 200 fs to 20 ps, in order to reduce the peak power and avoid non-linear self phase-modulation effects in the optical fiber used for coupling in the spectrometer. However, no relevant time information is lost due to the pulse expansion—because the time-resolved absorption is spectrally resolved, the dispersion can be reversed after the data is taken.

The time-resolved absorption is measured at $T$=70 K from PbSe nanocrystalline films deposited on fused silica substrates, shown in Figure 10a. The 1.3 μm output (FWHM = 17 nm) of the optical parametric oscillator was used both to pump and probe the system. Both the pump and the probe are intensity modulated using photo-elastic modulators at $f_1$ = 42 kHz and $f_2$ = 47 kHz respectively, and lock-in detection is performed at the difference frequency $f_R=f_1-f_2$ = 5 kHz. The resulting average time-resolved signal, spectrally averaged over the 320 wavelength channels of the lock-in spectrometer, is shown in Fig. 10b. The data collected over a similar time span using an identical experimental setup, but with a single InGaAs detector and conventional lock-in (Figure 10c) shows a deteriorated signal to noise ratio, arising from the less efficient data collection and the lack of demodulation capabilities along individual wavelength channels. The oscillations observed around zero delay for the lock-in spectrometer data arise due to the energy-time uncertainty relation and correspond to the 50 μeV spectral resolution of the spectrometer.

Time-resolved absorption experiments are also performed on self-assembled InAs quantum dots systems grown GaAs. The quantum dots are probed using a non-degenerate set-up, with the pump tuned to 775 nm and the probe tuned to 1240 nm. Carriers are created by the pump pulse in the GaAs, and relax quickly into the wetting layer and the quantum dots, where they eventually reach the ground state. Because the pump-probe experiment is non-degenerate, interference and scattering from the pump laser is eliminated, and it is possible to modulate the pump beam (at $f_R$=47 kHz) in order to provide the reference for phase-sensitive detection. Time-resolved absorption reveals long exciton decay times (> 1 ns) along all wavelength channels (Figure 11). The overall magnitude of the time-resolved signal is wavelength-dependent, being determined by the number of quantum dots that couple with the probe field at each wavelength. For the systems studied the quantum dot density is about $5 \times 10^{10}$ cm$^{-2}$, meaning that more than 200 dots are in the field of view of the in-cryostat 40x objective. Consequently, more than one quantum dot contributes to each wavelength channel and the envelope of the signal is uniform rather than discrete.

In summary, we have presented the theory and implementation of a multiple-channel lock-in optical spectrometer, capable of acquiring spectrally resolved data simultaneously over 320 wavelength channels, with a signal to noise ratio limited only by the laser noise, for intensity levels sufficient to overcome the intrinsic noise of the camera detector employed. The system implemented was optimized for the 1.3 μm wavelength range, although with a proper choice of acousto-optic modulator the performance of the system can be optimized for other wavelength ranges.

**Acknowledgments**

The authors are grateful to Gilberto Medeiros-Ribeiro for providing the InAs quantum dots samples, and to Yuichiro Kato for a careful reading of the manuscript. This work was supported by DARPA (DAAD19-01-1-0650).

**FIGURE CAPTIONS**

**Figure 1**  Schematic drawing of the lock-in spectrometer optical set - up. (FC) - fiber coupler; (GL) - Glan - Laser polarizer; (AO) - acusto- optical deflector; (L) - collimating lens; (DG) - diffraction grating part of a Czerny - Turner monochromator; (DSP) - digital signal processor.

**Figure 2**  Beam scanner alternatives: (a) Acousto - optical modulator; (b) Resonant scanner; (c) Polygon scanner.

**Figure 3**  Triggering diagram for the data acquisition and acousto-optical deflector control. The signals controlling the amplitude modulation are shown both for normal operation (──) and calibration (– – –).

**Figure 4**  Signal image extraction using the difference method to eliminate large backgrounds.

**Figure 5**  Graphic representation of the metric M extraction from the calibration signals for a single wavelength channel.

**Figure 6**  (a), (b) Images of the $\sin(2\pi f_R t)$ and $\cos(2\pi f_R t)$, respectively, reference vectors. (c), (d) line cuts along the wavelength channel 54.

**Figure 7**  Diode laser spectrum obtained after demodulation.

**Figure 8**  Noise dependence on the laser intensity

**Figure 9**  Optical set - up for a pump - probe experiment. (M) Mirror; (BS) Beam - splitter; (GLP) Glan-Laser polarizer; (OB) 40x objective; (FC) fiber coupler. The flipper mirror (FM) allows switching from a degenerate to a non - degenerate experiment.

**Figure 10** (a) Time – resolved absorption on PbSe films obtained by precipitation of colloidal nanocrystals; (b) Averaged data over the 320 channels of the lock - in spectrometer; (c) Data averaged over all the wavelengths using a InAs detector.

**Figure 11** Pump - probe time - resolved absorption on InAs quantum dots. Data acquired in parallel on 320 channels covering a wavelength range of 8 nm.

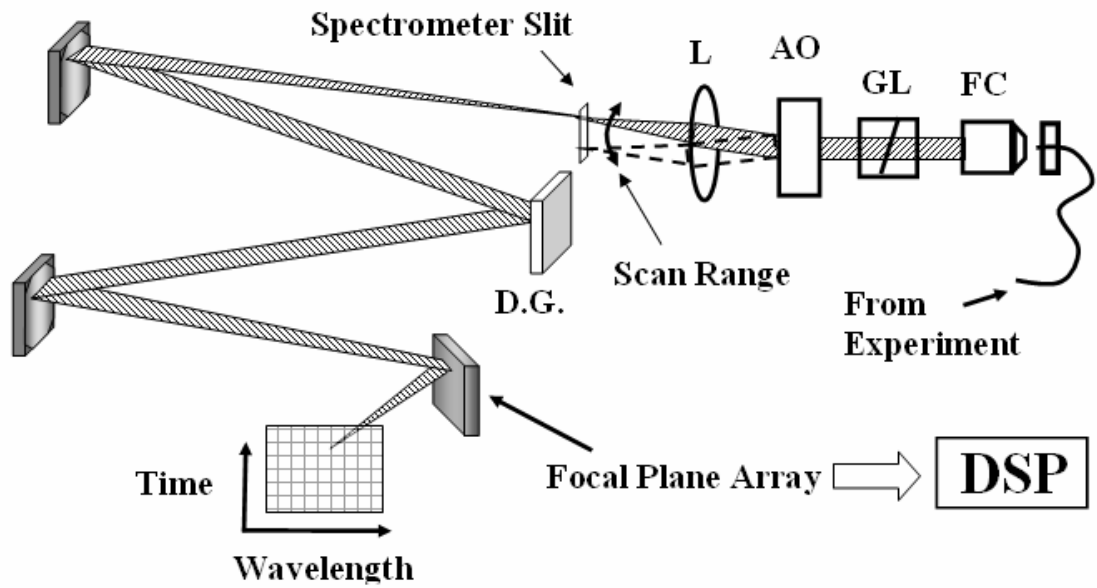

**Figure 1**  Schematic drawing of the lock-in spectrometer optical set - up. (FC) - fiber coupler; (GL) - Glan - Laser polarizer; (AO) - acusto- optical deflector; (L) - collimating lens; (DG) - diffraction grating part of a Czerny - Turner monochromator; (DSP) - digital signal processor.

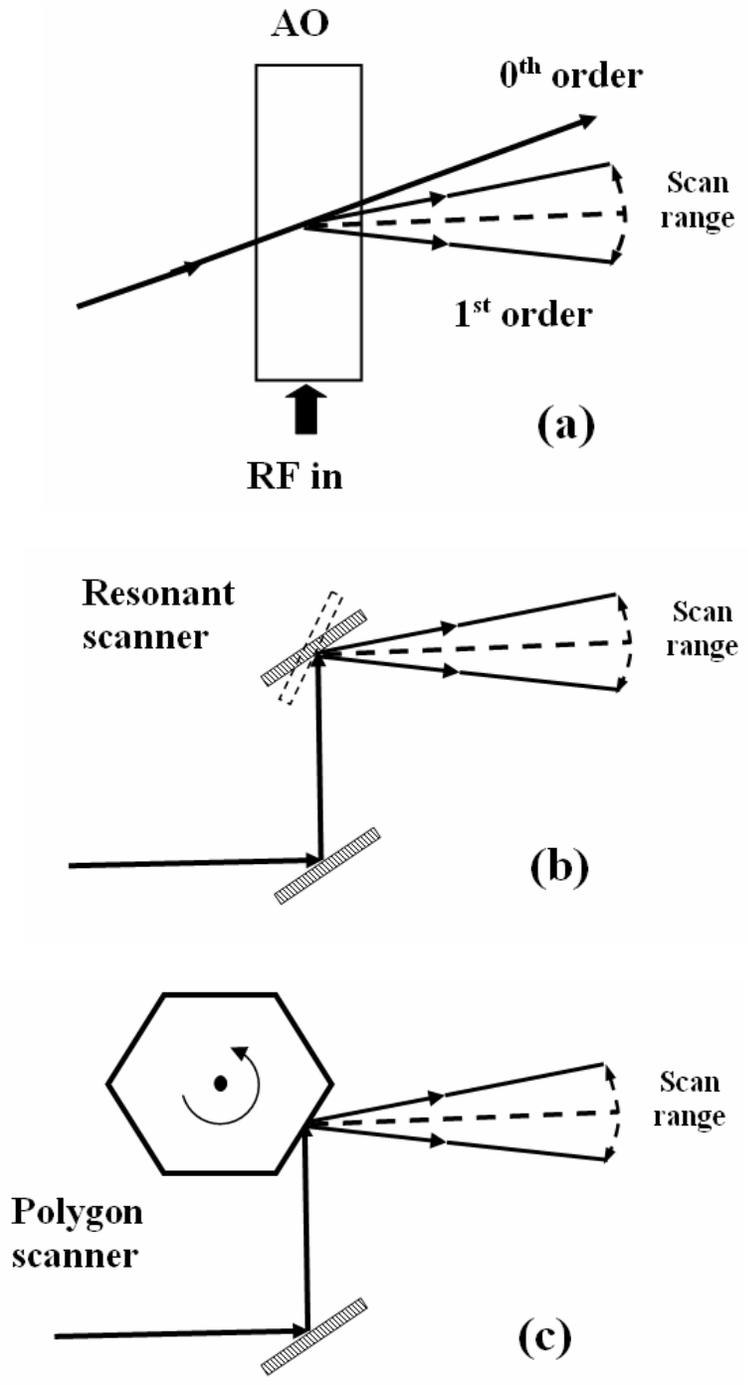

**Figure 2**  Beam scanner alternatives: (a) Acusto - optical modulator; (b) Resonant scanner; (c) Polygon scanner.

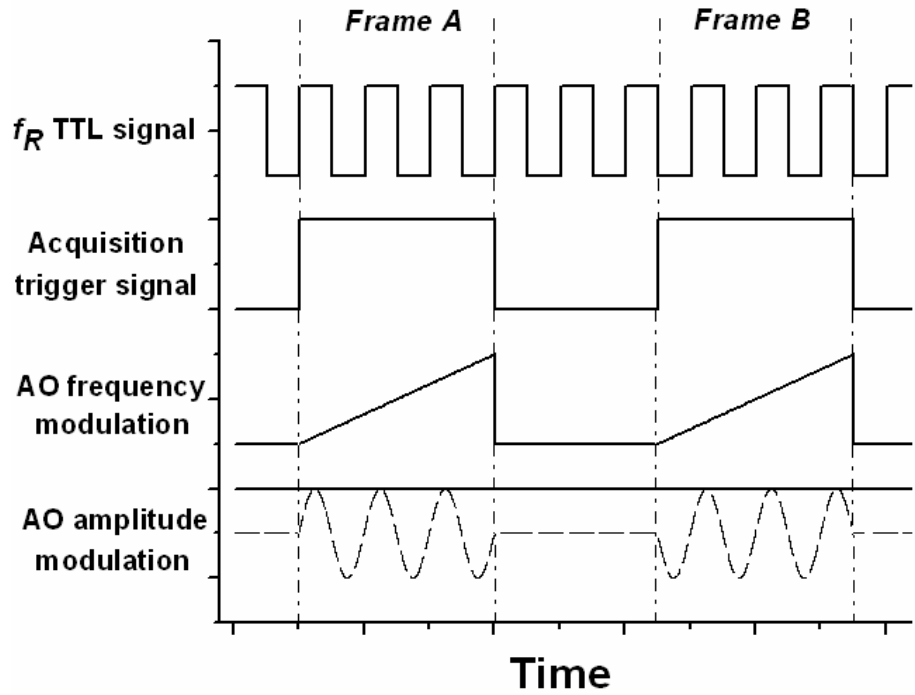

**Figure 3**    Triggering diagram for the data acquisition and acousto-optical deflector control. The signals controlling the amplitude modulation are shown both for normal operation (———) and calibration (– – –).

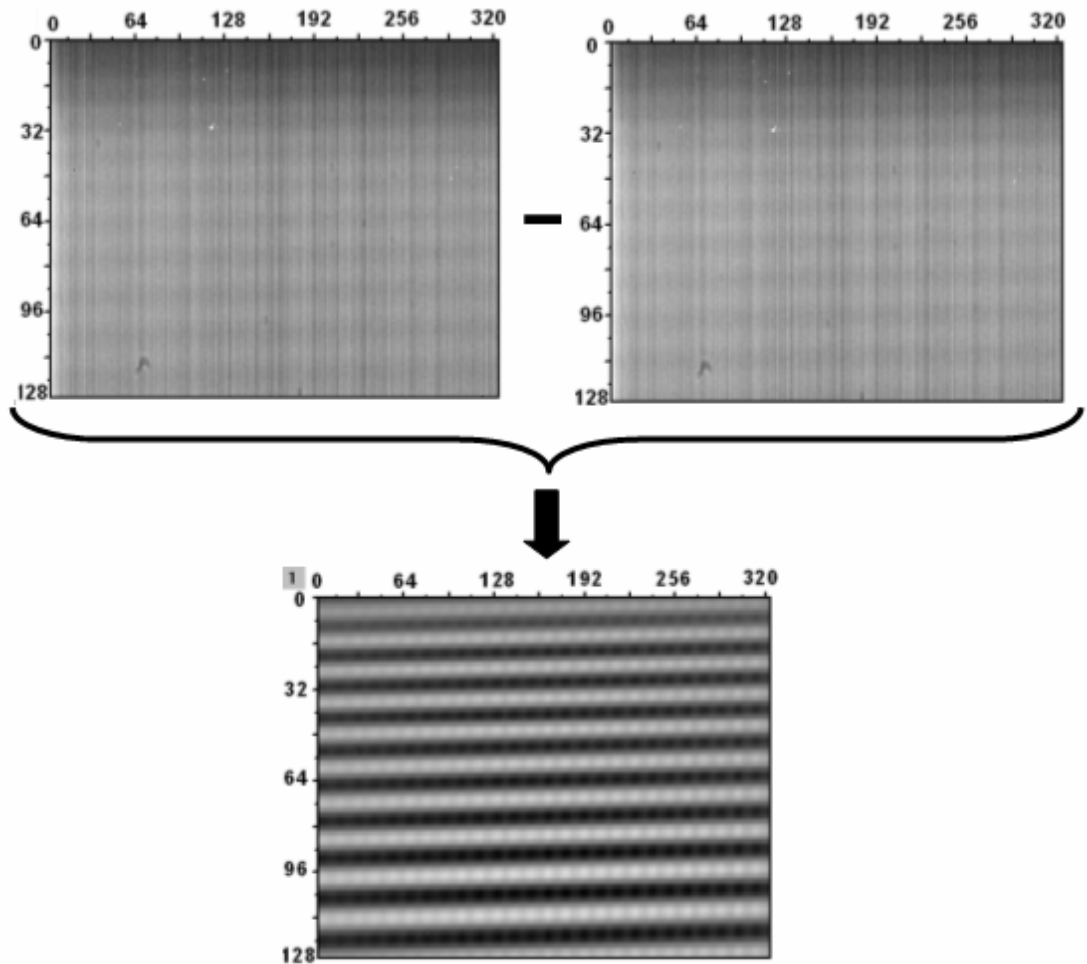

**Figure 4**   Signal image extraction using the difference method to eliminate large backgrounds.

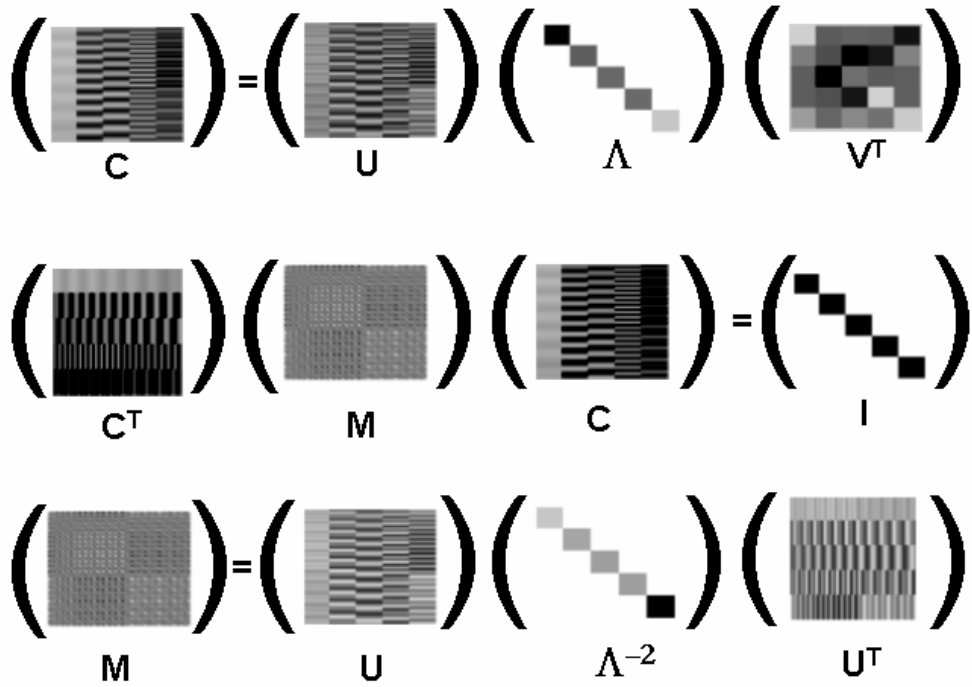

**Figure 5**  Graphic representation of the metric M extraction from the calibration signals for a single wavelength channel.

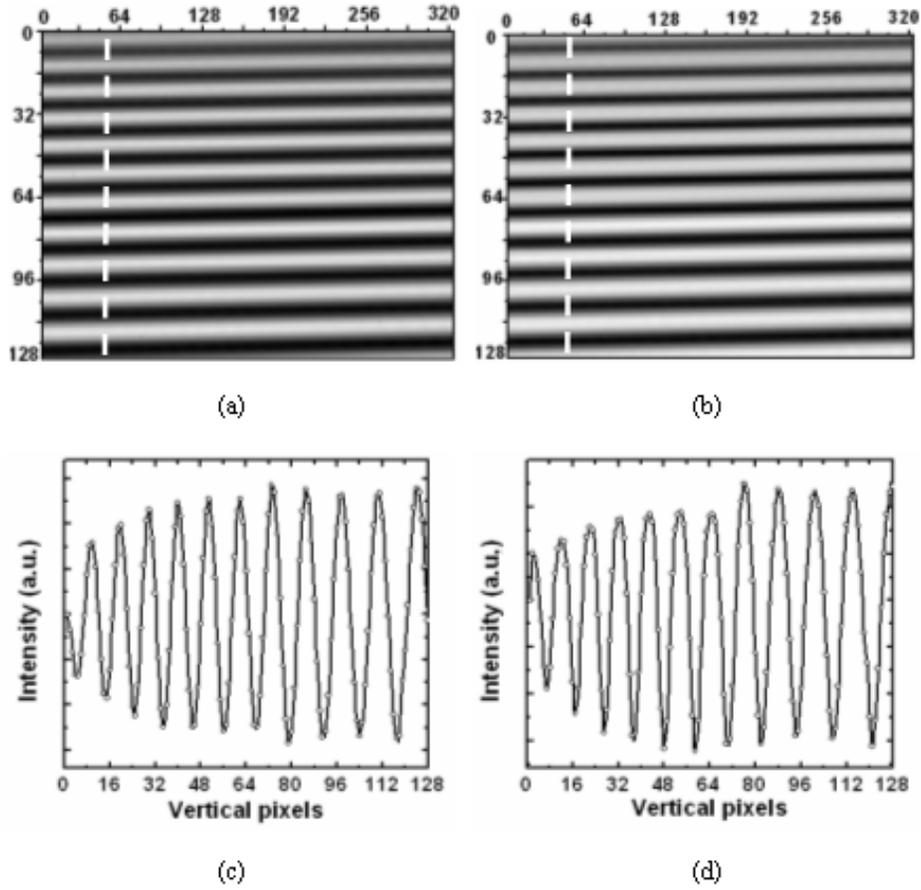

**Figure 6** (a), (b) Images of the $\sin(2\pi f_R t)$ and $\cos(2\pi f_R t)$, respectively, reference vectors. (c), (d) line cuts along the wavelength channel 54.

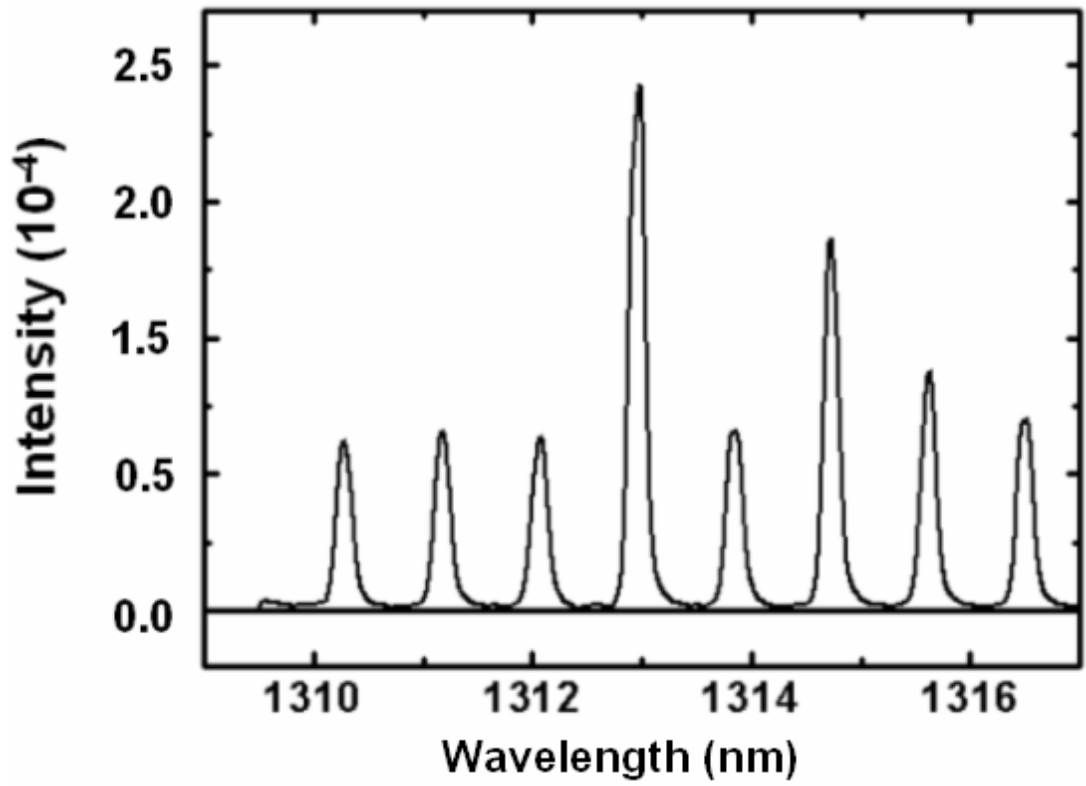

**Figure 7**     Diode laser spectrum obtained after demodulation.

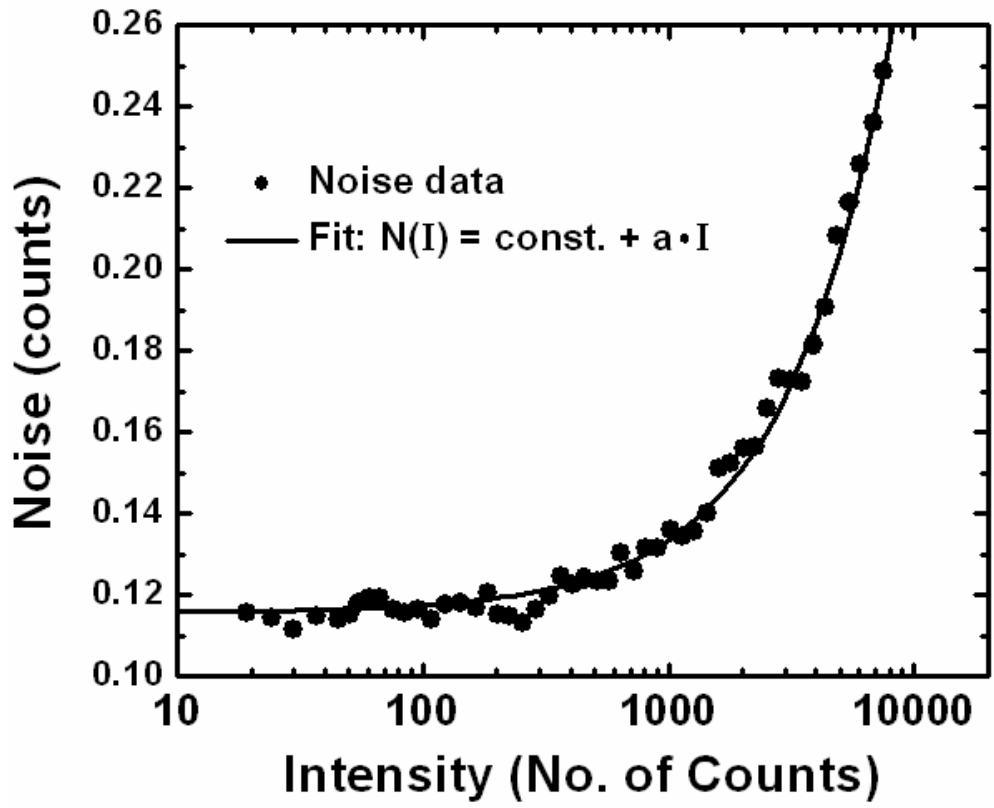

**Figure 8**    Noise dependence on the laser intensity.

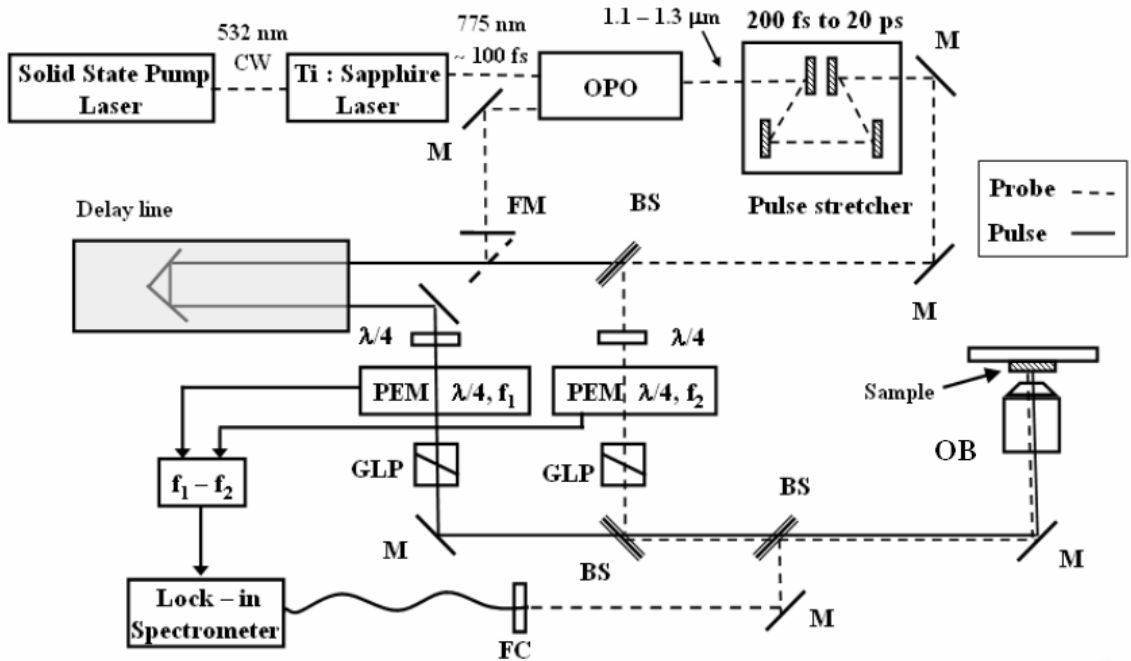

**Figure 9**  Optical set - up for a pump - probe experiment. (M) Mirror; (BS) Beam - splitter; (GLP) Glan-Laser polarizer; (OB) 40x objective; (FC) fiber coupler. The flipper mirror (FM) allows switching from a degenerate to a non - degenerate experiment.

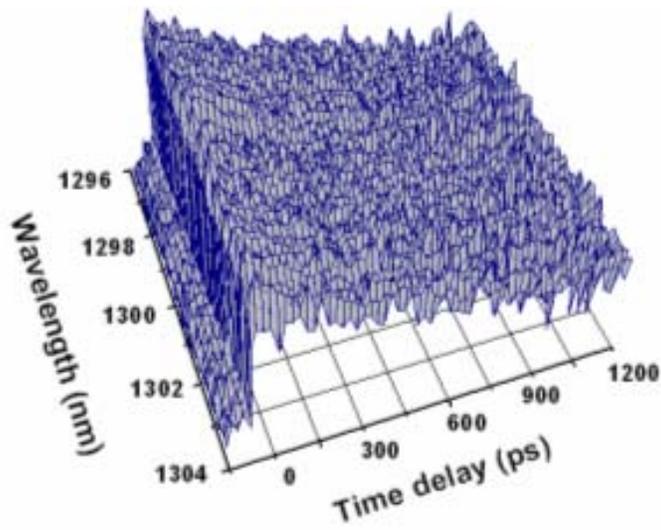

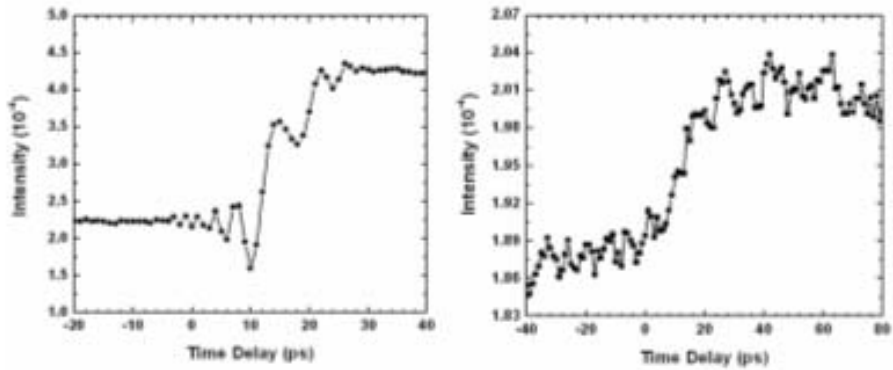

**Figure 10** (a) Time – resolved absorption on PbSe films obtained by precipitation of colloidal nanocrystals; (b) Averaged data over the 320 channels of the lock - in spectrometer; (c) Data averaged over all the wavelengths using a InAs detector.

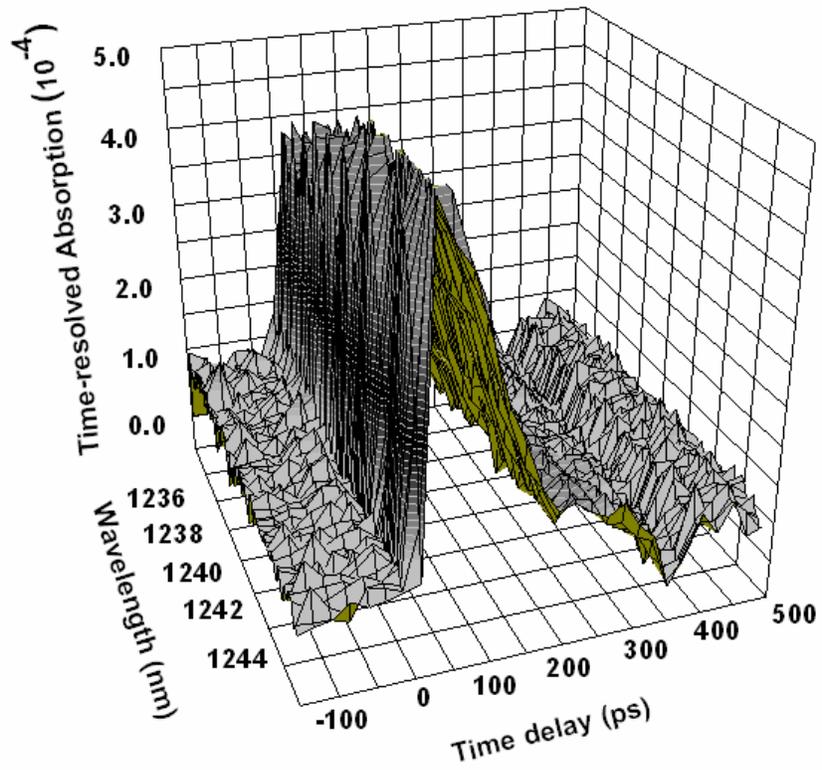

**Figure 11**  Pump - probe time – resolved absorption on InAs quantum dots. Data acquired in parallel on 320 channels covering a wavelength range of 8 nm.

27